\documentclass[%
 reprint,
superscriptaddress,
nofootinbib,
 amsmath,amssymb,
 aps,
 prd
]{revtex4-2}
\usepackage{multirow}
\pdfoutput=1
\usepackage{graphicx} % Required for inserting images
\usepackage[normalem]{ulem}
\usepackage[dvipsnames]{xcolor}
\definecolor{bluecite}{HTML}{0875b7}
\usepackage[colorlinks=true,urlcolor=Emerald,linkcolor=bluecite,citecolor=bluecite]{hyperref}
\usepackage[nameinlink]{cleveref}

\usepackage{amsmath,amssymb,amsfonts}
\usepackage[T1]{fontenc}
\usepackage[utf8]{inputenc}
\usepackage{bbm}

\newcommand{\hh}{\hat{h}}

\newcommand{\ie}{\textit{i.e.}}

\usepackage[nolist,nohyperlinks]{acronym}

\begin{document}

\title{\texorpdfstring{Comment on \\ \textit{Path integral measure and RG equations for gravity}}{Comment on \textit{Path integral measure and RG equations for gravity}}}

\author{Aaron Held}
\affiliation{Institut de Physique Th\'eorique Philippe Meyer, Laboratoire de Physique de l’\'Ecole normale sup\'erieure (ENS),
Universit\'e PSL, CNRS, Sorbonne Universit\'e, Universit\'e Paris Cit\'e, F-75005 Paris, France}

\author{Benjamin Knorr}
\affiliation{Institut f{\" u}r Theoretische Physik, Universit{\" a}t Heidelberg, Philosophenweg 16, 69120 Heidelberg, Germany}

\author{Jan M.~Pawlowski}
\affiliation{Institut f{\" u}r Theoretische Physik, Universit{\" a}t Heidelberg, Philosophenweg 16, 69120 Heidelberg, Germany}

\author{Alessia Platania}
\affiliation{Niels Bohr International Academy, The Niels Bohr Institute, Blegdamsvej 17, DK-2100 Copenhagen Ø, Denmark}

\author{Manuel Reichert}
\affiliation{Department  of  Physics  and  Astronomy,  University  of  Sussex,  Brighton,  BN1  9QH,  U.K.}

\author{Frank Saueressig}
\affiliation{High Energy Physics Department, Institute for Mathematics, Astrophysics, and Particle Physics, Radboud University, Nijmegen, The Netherlands}

\author{Marc Schiffer}
\affiliation{High Energy Physics Department, Institute for Mathematics, Astrophysics, and Particle Physics, Radboud University, Nijmegen, The Netherlands}

\begin{abstract}
Asymptotic safety is a candidate for a predictive quantum theory of gravity and matter. Recent works~\cite{Branchina:2024xzh, Branchina:2024lai} challenged this scenario. We show that their arguments fail on a basic level.
\end{abstract}

%---------------------------------------------------------------
\maketitle
%---------------------------------------------------------------

Asymptotic safety has been established as a 
candidate for the \ac{UV} closure of particle physics including gravity. The construction readily includes realistic sets of matter fields which makes the approach highly attractive from the phenomenological perspective.
Asymptotic safety builds on the modern perspective on renormalization: \ac{UV} complete \acp{QFT} are defined through fixed points of the \ac{RG} flow. Then, a \ac{QFT} of gravity would be defined via an interacting \ac{UV} fixed point. 

The development of functional \ac{RG} equations and lattice approaches for quantum gravity 
have put this scenario on a solid basis. In particular, by now overwhelming evidence has been accumulated for the existence of a suitable, diffeomorphism-invariant \ac{RG} fixed point, the Reuter fixed point. For introductory literature on this subject, see the textbooks~\cite{Percacci:2017fkn,Reuter:2019byg}; for a recent survey, see the handbook of quantum gravity~\cite{Bambi2024-nm}.

Recently,~\cite{Branchina:2024xzh, Branchina:2024lai} challenged this picture based on a non-standard perturbative one-loop computation. The essential claim is that the \ac{RG} fixed point underlying asymptotically safe quantum gravity disappears once the ``correct'' path integral measure is properly taken into account. In this note, we demonstrate that this conclusion is wrong for two key reasons:
\begin{enumerate}
\item[\textbf{(A)}] \textit{Fixed points define path integral measures:}

\Ac{UV}-complete \acp{QFT} are defined by \ac{UV}-stable fixed points. All functional \ac{RG} equations are capable of finding any fixed point, independently of the path integral measure. Fixing the measure and the action in the path integral as in~\cite{Branchina:2024xzh, Branchina:2024lai} restricts the fixed point search. Hence, one may miss physically relevant fixed points. This includes the Reuter fixed point, whose stability properties have been studied in abundance. An approach that tries to challenge it by restricting the search to a \textit{specific} measure \textit{and} action is ill-conceived from the start.

\item[\textbf{(B)}] \textit{Consistent flows are total scale derivatives:}

Fixed points exhibit scale invariance. Properly accounting for this requires the \ac{RG} equations to be a total derivative with respect to the coarse-graining scale. Only then points of vanishing flow are fixed points. 
The beta functions in~\cite{Branchina:2024xzh, Branchina:2024lai} are partial scale derivatives and their zeros do not define fixed points.

\end{enumerate}

Based on these points we conclude that the claims in~\cite{Branchina:2024xzh, Branchina:2024lai} with respect to the existence of the Reuter fixed point are incorrect. 
We also remark that, \textbf{(C)}, \textit{the one-loop divergences in~\cite{Branchina:2024xzh, Branchina:2024lai} do not agree with the standard ones}. Further discussions can be found in~\cite{Bonanno:2025xdg}. Below, we elaborate on the points \textbf{(A,B,C)}.

%---------------------------------------------------------------

%---------------------------------------------------------------
\subsection{Fixed points define path integral measures}
%---------------------------------------------------------------

In the path integral formulation for the metric fluctuations, the object of interest is 
\begin{align}
\int {\rm d}\mu\, e^{-S} \,,
\label{eq:PI}
\end{align}
where ${\rm d}\mu$ is the path integral measure and $S$ is an action functional. 
\eqref{eq:PI} entails that one actually should consider equivalence classes $(\textrm{d}\mu,S)$ with  
\begin{align} 
(\mathrm{d}\mu_1,S_1) \simeq  (\mathrm{d}\mu_2,S_2)\,,\quad \textrm{if}\quad \mathrm d\mu_1 \,e^{-S_1}=\mathrm d\mu_2 \,e^{-S_2}\,.
\label{eq:Pair}
\end{align}
Obviously, these pairs describe the same theory. Refs.\ \cite{Branchina:2024xzh, Branchina:2024lai} make a specific choice for both the measure and the action. Adopting this strategy may result in an ill-defined theory. This can be seen as follows. Consider scalar $\varphi^4$-theory in three dimensions with a $\mathbbm{Z}_2$ symmetry. Then the pair $(\mathrm{d}\mu_1, S_1)$ given by the flat measure, $\mathrm d\mu_1 = \mathrm d\varphi$, and the classical action
\begin{equation}
 S_1[\varphi] = \int_x \bigg\{ \frac{1}{2} (\partial_\mu\varphi)^2 +\frac{m^2}{2} \,  \varphi^2 +\frac{\lambda_4}{4!}\, \varphi^4\bigg\} \, ,
\end{equation}
defines a super-renormalizable theory that is asymptotically free. The fixed point action is the free scalar action. It has two \ac{UV}-relevant directions, the mass $m$ and the coupling $\lambda_4$. The theory also features an attractive infrared fixed point, the Wilson-Fisher fixed point. This fixed point is reached from the \ac{UV} for a unique choice of the \ac{UV}-relevant parameters in the $\varphi^4$ theory. 

Consider now the following change of the theory: we change the measure to
\begin{align} 
\mathrm d\mu_2[\varphi] = \mathrm d\varphi \, e^{-\int_x\lambda_8\,\varphi^8}\,, 
\end{align}
while keeping the action fixed, \ie, we consider the pair $(\mathrm{d}\mu_2,S_1)$. This pair is equivalent to the classical action $S_2$ of a $\varphi^8$ theory, $S_2=S_1 +\int_x \lambda_8\,\varphi^8$ with the flat measure, that is $(\mathrm{d}\mu_2,S_1) \simeq (\mathrm{d}\varphi, S_2)$. The coupling $\lambda_8$ has a negative mass dimension, and the resulting theory is non-renormalizable. The theory breaks down in the \ac{UV} and no path integral exists for the pair $(\mathrm d\mu_2,S_1)$. 

Evidently, the inconsistency of the pair $(\mathrm{d}\mu_2, S_1)$, does not invalidate the existence of an asymptotically free scalar field theory in three dimensions.\\[-1.25ex]

The Wilsonian \ac{RG} defines a quantum theory by the flow equation of its generating functional. A specific example is the Wetterich equation for the one-particle irreducible effective action $\Gamma_k$,
\begin{equation}
    k \partial_k \Gamma_k = \frac{1}{2} \text{Tr} \left[ \left( \Gamma_k^{(2)} + R_k \right)^{-1} k \partial_k R_k \right] \, .
    \label{eq:flow}
\end{equation}
Importantly, all of these flows only depend on the generating functional at hand and its derivatives. In particular, they depend neither on the classical action nor the measure.
Comparing \eqref{eq:flow} with the path integral \eqref{eq:PI} entails that a \ac{QFT} can either be defined by the \ac{UV} fixed point action $\Gamma_\textrm{FP}$ or, equivalently, by a pair \eqref{eq:Pair}. Hence, the existence of a $\Gamma_\textrm{FP}$ implies the existence of a specific equivalence class $(\mathrm d\mu,S)=(\mathrm d\mu,S)[\Gamma_\textrm{FP}]$ for which the path integral is well-defined in the \ac{UV}. \\[-1.25ex]

This leaves us with the following conclusion for asymptotic safety: The Reuter fixed point has been established as a diffeomorphism-invariant, \ac{UV}-stable fixed point with only a few relevant directions and close-to-canonical scaling for irrelevant operators. These latter properties facilitate the stability and systematic error analysis. The respective systematic analyses have been performed over the past 30 years and by now its existence (in pure gravity) is very well-established. In short, the measure discussion in~\cite{Branchina:2024lai, Branchina:2024xzh} is irrelevant when assessing the existence of this fixed point.

%--------------------------------------------------------------
\subsection{Consistent flows are total scale derivatives} 
%--------------------------------------------------------------

Fixed points are points of vanishing flow with respect to total, not just partial, derivatives of the \ac{RG} scale $k$. The derivation in~\cite{Branchina:2024xzh, Branchina:2024lai} violates this principle of the fixed-point search by an unconventional cutoff identification and the subsequent neglect of implicit scale dependence. 

We highlight the key steps. The works~\cite{Branchina:2024xzh, Branchina:2024lai} introduce powers of the background curvature $R=12/a^2$ as the reference scale in the computation, in combination with a redefinition of the quantum fields (see~\cite[eq.~(11)]{Branchina:2024xzh}),
\begin{equation}\label{field-redef}
\hh_{\mu\nu} \equiv (32 \pi G)^{-1/2} \, a^{-1} \, h_{\mu\nu} \, .
\end{equation} 
Implicit scale dependence needs to be taken into account carefully when redefining the fundamental fields and/or the cutoff identification.
As highlighted in~\cite{Branchina:2024xzh, Branchina:2024lai} (see~\cite[Sec.~3]{Branchina:2024xzh} as well as~\cite[Sec.~5]{Branchina:2024lai}), the standard literature relates the cutoff
\begin{align}
    \Lambda_{\rm cut} = \frac{N}{a}\,,
    \qquad
    \text{(standard)}
\end{align}
to $N$, which enumerates the eigenvalues of the Laplacian, and the radius $a$ of the background/reference metric. By contrast, \cite{Branchina:2024xzh, Branchina:2024lai} specify the cutoff
\begin{align}
    \Lambda_{\rm cut} = \frac{N}{a_{\rm dS}(k)}\,,
    \qquad
    \text{(\cite{Branchina:2024xzh, Branchina:2024lai})}
\end{align}
in terms of the $k$-dependent \emph{on-shell} radius $a_{\rm dS}$ of the four-dimensional sphere.

While $a$ is a background quantity, \ie{}, it does not depend on the \ac{RG} scale, $a_{\rm dS}$ is an on-shell quantity which itself depends on the running couplings (see~\cite[eq.~(40)]{Branchina:2024xzh}) and hence carries an implicit scale-dependence: $a_{\rm dS}=a_{\rm dS}(k)$. \cite{Branchina:2024xzh, Branchina:2024lai} do not account for this implicit scale dependence. Hence, their flow does not correspond to a total derivative of the \ac{RG} scale $k$.\\[-1.25ex]

In conclusion, the zeros of the corresponding equations do not define fixed points.

%---------------------------------------------------------------
\subsection{Non-standard one-loop divergences}
%---------------------------------------------------------------

The field redefinitions in~\cite{Branchina:2024xzh, Branchina:2024lai} make all fields and operators dimensionless. This changes the divergence structure of loop diagrams. We illustrate this feature based on the ghost contribution to the graviton two-point function in flat space. There, standard one-loop \ac{QFT} methods lead to three types of divergences, namely $p^4 \log(\Lambda_{\mathrm{UV}})$, $p^2\Lambda_{\mathrm{UV}}^2$, and $\Lambda_{\mathrm{UV}}^4$, where $p$ is the external momentum and $\Lambda_{\mathrm{UV}}$ is the \ac{UV} cutoff. By diffeomorphism invariance, these renormalize the $R^2$-coupling, the Newton coupling, and the cosmological constant, respectively. On the other hand, \cite{Branchina:2024xzh, Branchina:2024lai} only find a $p^4 F(N)$ divergence, where $F(N)$ is some function of the numerical cutoff $N$. On a spherical background, this term entails a renormalization of the $R^2$-term only and not the other two operators, in clear contradiction to the standard computation. The reason for this shift of divergences into only the $R^2$-term is the transition from the metric $\bar{g}$ on a sphere of radius $a$ to the metric $\tilde{g}$ on a unit sphere in~\cite{Branchina:2024xzh, Branchina:2024lai}. This is essential for the dimensionless nature of the differential operators, and results in a non-standard definition of variational derivatives, which now have to be normalized with $\sqrt{\det(\tilde{g})}$. Similar arguments apply to the gravitational sector.

More concretely, the $p^4$-term comes from measuring operators in units of the background curvature, while setting $\tilde{R}=12$. This means that the only scale, namely the background curvature, is used to make operators dimensionless. As a result, different monomials $\tilde{R}^n$ cannot be distinguished, such that disentangling the flow of different curvature operators is impossible. Thus, the calculations in~\cite{Branchina:2024xzh, Branchina:2024lai} are in tension with standard one-loop computations.

%---------------------------------------------------------------
\subsection*{Acknowledgements}
We thank A.\ Bonanno, A.\ Eichhorn, G.\ de Brito, K.\ Falls, R.\ Ferrero, D.\ Litim and R.\ Percacci for discussions and comments on the manuscript. \vfill

\begin{acronym}[EFT]
\acro{UV}[UV]{ultraviolet}
\acro{QFT}[QFT]{quantum field theory}
\acrodefplural{QFT}{quantum field theories}
\acro{RG}[RG]{renormalization group}
\end{acronym}

\bibliography{bibliography}

\end{document}